\documentclass[12pt,a4paper]{article}

\usepackage[textheight=23cm,textwidth=16cm]{geometry}
\usepackage{amsmath}
\usepackage{siunitx}
\usepackage{appendix}
\usepackage{tabularx}
\usepackage{graphicx}
\usepackage{amssymb}
\usepackage[american]{babel}
\usepackage{cite}
	\newlength\myboxwidth	
\usepackage{multirow}
\usepackage[utf8]{inputenc}

\usepackage[table, dvipsnames]{xcolor}
\usepackage{soul}

\usepackage{amsfonts}

\usepackage[hidelinks]{hyperref}

\usepackage{url}
\usepackage{xcolor}
\usepackage{lineno}

\usepackage[affil-it]{authblk}

\usepackage{xspace}
\begin{document}

\renewcommand\thesection{\Roman{section}}
\renewcommand\thesubsection{\arabic{subsection}}
\renewcommand{\labelitemii}{$\longrightarrow$}

\begin{center}

{\LARGE\bf
 Universal QM/MM Approaches for General Nanoscale Applications \\}

\vspace{1cm}

\renewcommand*{\thefootnote}{\fnsymbol{footnote}}

{\large
Katja-Sophia Csizi and Markus Reiher\footnote{Corresponding author; e-mail: markus.reiher@phys.chem.ethz.ch}
}\\[4ex]

\renewcommand*{\thefootnote}{\arabic{footnote}}
\setcounter{footnote}{0}

ETH Z{\"u}rich, Laboratorium f\"ur Physikalische Chemie, Vladimir-Prelog-Weg 2,\\ 
8093 Z{\"u}rich, Switzerland 

\vspace{.5cm}

December 19, 2022

\vspace{.43cm}

\textbf{Abstract} 
\end{center}
\vspace{-0.5cm}
  
{\small Quantum mechanics/molecular mechanics (QM/MM) hybrid models allow one to address chemical phenomena in complex molecular environments. Whereas this modeling approach can cope with a large system size at moderate computational costs, the models are often tedious to construct and require manual preprocessing and expertise. As a result, transferability to new application areas can be limited and the many parameters are not easy to adjust to reference data that are typically scarce. Therefore, it is desirable to devise automated procedures of controllable accuracy, which enables such modeling in a standardized and black-box-type manner. Although diverse best-practice protocols have been set up for the construction of individual components of a QM/MM model (e.g., the MM potential, the type of embedding, the choice of the QM region), automated procedures that reconcile all steps of the QM/MM model construction are still rare. Here, we review the state of the art of QM/MM modeling with a focus on automation. We elaborate on MM model parametrization, on atom-economical physically-motivated QM region selection, and on embedding schemes that incorporate mutual polarization as critical components of the QM/MM model. In view of the broad scope of the field, we mostly restrict the discussion to methodologies that build \textit{de novo} models based on first-principles data, on uncertainty quantification, and on error mitigation with a high potential for automation. 
Ultimately, it is desirable to be able to set up reliable QM/MM models in a fast and efficient automated way without being constrained by specific chemical or technical limitations.
}

\newpage
\section{Introduction}
The exploration of chemical structure space and its corresponding reaction space for nanoscopic atomistic systems requires computational models that account for both their physically consistent description and their size, comprising several thousand atoms. 
Multiscale approaches have emerged as a powerful methodology to tackle this task, whose central idea is a combination of quantum mechanics (QM) and classical molecular mechanics (MM).\cite{Warshel2013} 
Conceptually, this merger is achieved by the simple idea to marry these two different mechanical theories at the level of the total energy of a system, which defines the QM/MM modeling approach~\cite{Warshel1976}. 

The reliable description of structural changes and chemical reactions requires an electronic structure method to produce reliable sections of the potential energy surface (PES) of an atomistic system in the Born--Oppenheimer approximation. However, electronic structure methods scale with some power of a measure for system size so that their applicability is limited to comparatively small systems only, that is, to those that are usually much smaller than nanoscopic ones. Since breaking or forming chemical bonds are local processes, the idea of a QM/MM separation with a comparatively small QM region can account for them. 

The QM/MM hybrid approach separates a molecular system into a quantum mechanically described core region $\mathcal{Q}$, to which empirical valence bond theory~\cite{Warshel1980, Kamerlin2010, Kamerlin2011}, semiempirical methods~\cite{Govender2014}, and first-principles approaches~\cite{Torabifard2018a, Nicholson2020} can be applied, and an environment region $\mathcal{E}$ modeled with a classical MM force field (FF) in such a way that the total energy is obtained as
\begin{equation}
	E_{QM/MM} = E_{QM}^{\mathcal{Q}} + E_{MM}^{\mathcal{E}} + E_{int}^{\mathcal{Q} + \mathcal{E}}
	\label{eq:qmmm-energy}
\end{equation}
with $E_{QM}^{\mathcal{Q}}$ being the electronic energy of the quantum subsystem $\mathcal{Q}$, $ E_{MM}^{\mathcal{E}}$ the energy of the environment subsystem $\mathcal{E}$, and $E_{int}^{\mathcal{Q} + \mathcal{E}}$ the potential energy of the interaction of both subsystems. 

This setting introduces a quantum magnifying glass on a local region in which electronic processes dominate, such as chemical reactions and photo excitations~(see Refs. \citenum{Lin2006, Senn2007, Senn2009, Orlando2010, Wallrapp2011, Akimov2015, Brunk2015, Doll2015, Quesne2016, Zheng2016a, Hofer2018, Magalhaes2020, Lipparini2021} for recent reviews). Typical areas of application are biomolecules~\cite{Sousa2017}, metalloenzymes~\cite{Benediktsson2017, Nicholson2020}, symmetry defects in metals~\cite{Huber2016}, covalent~\cite{Cui2020} and metal~\cite{Moeljadi2016} organic frameworks, carbon-nanotubes~\cite{Khare2007} and graphene~\cite{Restuccia2020}, solution chemistry in studies of spectroscopy~\cite{Morzan2018}, light induced processes~\cite{Mennucci2019, Boulanger2018}, solvation ~\cite{Pezeshki2014a, Zhang2018c, Boereboom2018, Clabaut2020a} and adsorption~\cite{Clabaut2020a} chemistry, excited states~\cite{Marefat2020}, and nonadiabatic dynamics~\cite{Avagliano2021}. Also, quantum classical hybrid models have been leveraged to accelerate explicitly atomistic drug design efforts~\cite{Koenig2021, Sabe2021}, with the aim to improve on the accuracy of classical MM models and to cover nonstandard chemistry by a QM approach.\cite{Kar2023}
The QM-cluster approach\cite{Siegbahn2011,Ahmadi2018} is an alternative approach to deal with a large nanoscopic system. Structural constraints can be considered in these models by frozen atomic positions in the periphery, but the general mechanical behavior of, for instance, a protein will be difficult to capture. Accordingly, QM-cluster models have been well established for solvation phenomena~\cite{Kirchner2007}, but convergence of a physical quantity with cluster size can be difficult to achieve.\cite{Simm2020, Bensberg2022}

Besides the total energy, analytic first derivatives thereof with respect to the nuclear coordinates are mandatory for the efficient optimization or propagation of molecular structures and for the exploration of a PES.
The evaluation of analytic second derivatives, that is, of the Hessian matrix, is computationally very expensive, because additional coupled perturbed self-consistent field equations may need to be solved for each MM atom displacement~\cite{Cui2000, Ghysels2011}.
Naturally, approximate strategies have been devised such as neglecting the MM degrees of freedom in a partial Hessian approach~\cite{Head1997}, approximating the coupled perturbed self-consistent field equations~\cite{Cui2000, Ghysels2011}, exploiting a fragment molecular orbital ansatz~\cite{Nakata2013}, the partial Hessian vibrational approach~\cite{Giovannini2019}, and the Q-vector method~\cite{Schwinn2019, Schwinn2020}.

Although multiscale hybrid models can, in principle, be applied to any atomistic system, this is often hampered by the lack of suitable QM/MM models. For instance, for chemical reactions on molecular surfaces, it is difficult to model bulk properties in a QM calculation and across the QM/MM boundary. 

The sheer size of nanoscopic structures composed of several thousands of atoms requires enormous manual input in the construction of both the atomistic structure and the corresponding QM/MM model. 
Therefore, QM/MM methods exhibit technical and methodological difficulties which need to be carefully considered.\cite{Lin2006, Senn2007, Yang2010, Lu2016, Magalhaes2020, Cui2021} 
These difficulties may be classified roughly according to:
\begin{enumerate}
\item Atomistic structure generation.
\item Choice and spatial size of the QM region.
\item 
Choice and parametrization of the analytic terms in the FF.
\item Embedding and handling of the QM/MM interface.
\item Reproducibility and automation.
\item Uncertainty quantification (UQ)/Error control.
\end{enumerate}
Consequently, the number of tunable parameters and the choices that need to be made during the construction of a QM/MM model open a pandora's box of decisions to be taken that strongly affects the outcome and reliability of a QM/MM simulation.
To mitigate these issues, we define a set of target properties required for the construction of a flexible QM/MM model that can be adopted to any nanoscopic system: 
\begin{enumerate}
\item First-principles-based parametrization of the classical part (as only such a feature will allow one to produce electronic-structure reference data on demand so that a lack of data does not pose a problem for model construction).
\item Rigorous determination of the QM region by physical criteria with (asymptotically) converged properties while retaining computational feasibility (because only this property will allow for automated model construction and guarantee a seamless QM embedding with a handle on model error assessment).
\item Physically meaningful description of the subsystem interface (in order to avoid or reduce model artifacts).
\item UQ for model parameters and simulation (to allow for model intrinsic error assessment and to increase predictive power and transferability).
\item Agnosticism with respect to the elemental composition of a composite nanoscopic system (since a restriction toward certain elements of the periodic table would pose a severe limitation toward general nanoscale applicability).
\item Algorithmic robustness (in order to not corrupt automation and high throughput simulation).
\item Favorable cost-to-accuracy ratio (which is a moving target as it depends on various factors and therefore requires automated procedures in all model construction steps in order to adjust model accuracy seamlessly when needed).
\end{enumerate}
In this work, we focus on development aspects of QM/MM models that fulfill these properties toward generally applicable QM/MM models. Note that we omit the broad topic of free energy calculations~\cite{Christ2010, Hansen2014, Lu2016, Pietrucci2017, Armacost2020}, advanced sampling techniques~\cite{Spiwok2015, Invernizzi2020, Kamenik2022}, and automated reaction network exploration~\cite{Sameera2016, Dewyer2017, Dewyer2018, Vazquez2018, Simm2019, Unsleber2020, Maeda2021, Steiner2021, Unsleber2022a, Baiardi2022a},
which can deal with the huge size of configuration space and chemical space,
as this is beyond the scope of this work.
First, we briefly discuss the aforementioned shortcomings in the setup of a multiscale model and then present state-of-the-art solutions with a reference to applications in order to highlight the progress achieved so far.

\section{The Need for Automatically Constructible QM/MM Models}
The amount of tunable adjustment knobs and, by that, the significant manual work involved in the generation of a QM/MM model underlines the need for automated protocols for model construction that are then applicable to nanoscopic systems of arbitrary elemental composition. 
However, the development of methodologies that automatically assemble such convergent QM/MM models appears to have not kept pace with the rate of application of QM/MM methodology. 

Automated protocols offer two main advantages: First,
applications become feasible even for those who are nonexperts in the field, that is, for those to whom a manual QM/MM approach would be new or inaccessible. 
Second, a black-box first-principles-based model generator allows for the study of molecular systems in an unbiased way, because no constraints are imposed by either chemical intuition or the presence (or absence) of experimental data. Moreover, such a generator would ensure that structure-, force-, and energy-wise the MM model resembles the electronic structure model as closely as possible and therefore clearly separates electronic and nuclear contributions (note that this is a quantum chemistry view of MM modeling and other ways to parametrize MM models for different foci (e.g., reproduction of thermodynamical data\cite{vanGunsteren2006}) exist as well.

Even further, as it is desirable to establish procedures for the systematic construction of hundreds of QM/MM models for diverse applications or automated screening in high-throughput explorations, only an autonomous parametrization scheme that can rely on reliable training data produced on demand can deliver this. 

Although automatized procedures are available for the setup of either one of the QM/MM components (FF parametrization, QM region determination, structure preparation and embedding), an approach that combines automated protocols assembling all components in a full-fledged black-box QM/MM model generator has been lacking. Recently, our group introduced such an autonomous QM/MM model generation protocol that combines system-focused, quantum-chemical FF parametrization to obtain a system-focused atomistic model (SFAM)~\cite{Brunken2020} coupled to a completely automated set-up for a QM/MM hybrid model (QM/SFAM)~\cite{Brunken2021}. The QM/SFAM model (and by that, the molecular structure) can be iteratively refined by on-the-fly generated reference data and a machine learning (ML) approach. 
A schematic workflow of this scheme is depicted in Figure 1.  

\clearpage
\begin{figure}[htp!]
\includegraphics[width=1.0\textwidth]{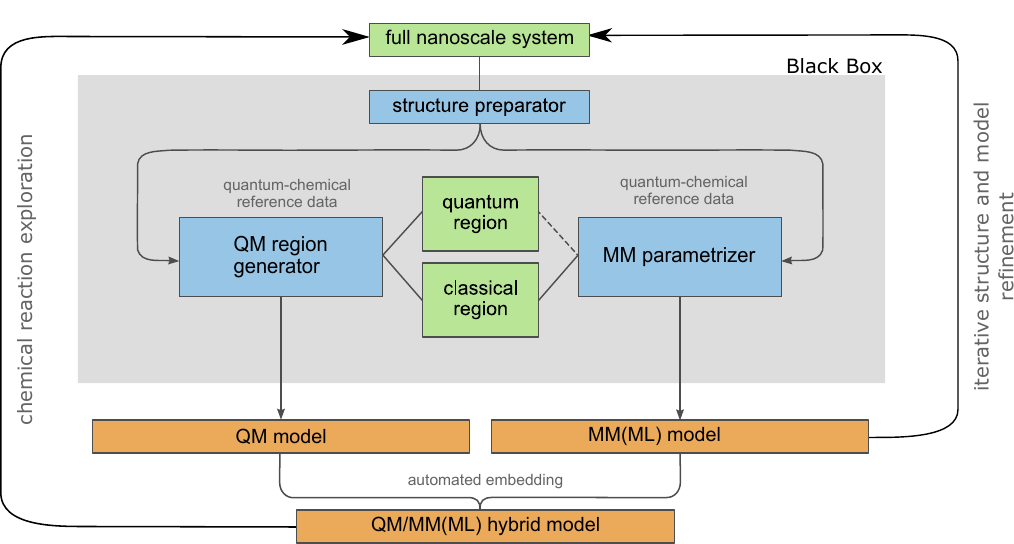}
\label{fig:automation}
\caption{A workflow for the construction of the fully-automated QM/MM SFAM protocol following Ref. \cite{Brunken2021}.}
\end{figure}

\section{Structure Generation}
Structure and function in nanoscopic systems are tightly coupled.  
For instance, the reaction specificity of enzymes is determined and modulated by the structure of the active site pocket. 
Despite advances in data collection and resolution, experimentally determined structures are still subject to significant uncertainties in atomic positions and suffer from multiple structural errors~\cite{Davis2003,Davis2007} such as unreasonable geometries, atomic clashes, or incompletely resolved residues.
These issues negatively affect, among others, binding affinity prediction~\cite{Davis2003} and drug-design efforts~\cite{Warren2012, Borbulevych2018}. 
Therefore, structures determined by experiment must be carefully analyzed and possibly refined before being subjected to QM/MM simulations (since every atom and its specific position will have a decisive effect on the electronic structure and hence the energy of the whole system). This can be done by built-in workflows for analyzing and clean-up of corrupted experimental raw data for molecular structures.\cite{Melaccio2016a, Pedraza-Gonzalez2019, Pedraza-Gonzalez2021} 
We note that various pipelines are available that help in the structure preparation process for atomistic simulations (see, for instance, Refs. \citenum{Dolinsky2007, Jo2008, Anandakrishnan2012}), but they are mostly restricted to biomolecular applications.

Even if a sufficiently accurate crystal structure is available, it may still not be suited for QM/MM simulations. 
The reasons for this are: (i) X-Ray diffraction structures generally depict an early stage of enzyme catalysis due to the fact that the enzymatic cycle proceeds fast and intermediates are not always resolved (or only under manipulation, e.g., at low temperature or structural analogues of substrates or through time-resolved imaging applying free electron lasers~\cite{Aquila2012}). 
(ii) X-Ray diffraction structures contain ensemble information on alternate conformations which are deposited in a single crystal structure, among which multiple or all may need to be considered. 
(iii) Hydrogen atoms are usually not resolved in X-Ray diffraction experiments (but need to be studied with special techniques such as neutron scattering). Information about them needs to be recovered, not only because enzymatic function and activity is closely regulated by the physiological environment and its pH value~\cite{Bergdoll2017, Sun2017}, which determines the protonation state of pH-sensitive groups, but any electronic structure calculation that serves for training data generation of a FF and for the QM region will crucially depend on every individual atom and electron (none should be missed). 

Other experimental techniques than X-ray diffraction are also used for structure elucidation (most prominently, nuclear magnetic resonance spectroscopy), but they also come with specific issues regarding time resolution and ensemble averaging.

Conventional crystal structure preparation and refinement is usually based on classical FFs, where a raw molecular structure is subjected to an MD simulation, from which appropriate snapshots can be fed into QM/MM modeling afterward. 
Both, number~\cite{Ryde2017} and choice~\cite{Esch2019} of snapshots impose some constraints on which region of the  PES will be visited during the simulation. This selection can be guided by many different factors such as the intrinsic autocorrelation time of the process of interest~\cite{Chodera2016}, the phase space overlap between the MM and QM potential energy surfaces~\cite{Koenig2018a}, or the choice of approximations and constraints.\cite{Hudson2019}

Since available FFs are not equipped with accurate parameters for nonstandard chemistry, this fact facilitated the development of quantum refinement methods. 
The basic idea goes back to work of Ryde and coworkers~\cite{Ryde2002} and is in the  spirit of the QM/MM method itself, because critical regions, such as metal sites, can be treated quantum mechanically~\cite{Yan2021}. In a QM/MM refinement loop, the raw input structure is iteratively optimized until convergence of a specific property is achieved (for example, atomic charges, structural changes, the overall energy). Application areas of this approach have been, for instance, the determination of protonation states~\cite{Borbulevych2018, Borbulevych2021a}, even for poorly resolved structures~\cite{Borbulevych2016}, and the differentiation between different bonding patterns in active sites of metalloenzymes.\cite{Cao2018a} 

Within our SFAM and QM/SFAM approaches~\cite{Brunken2020,Brunken2021}, structure optimizations can be performed on the initial structure with an automatically parametrized model. This is then followed by iterative model and structure refinement. 
Such a feedback-loop not only reduces uncertainty in the structure, but also allows one to construct nanoscale models bottom-up for systems for which no extensive structure information is available (by contrast to protein structures deposited in the protein data bank). 

\section{Definition of a Suitable Force Field}

\label{sec:force-fields}
By their very nature as approximate potential energy functions, parametrized
FFs are usually available only for certain classes of molecules, but not for a specific or general molecular system that does not belong to any of these classes. 
Instead, one either picks a universal transferable FF (typically at the cost of reduced reliability) or apply (re-)parametrization procedures to obtain a tailored system-specific potential. 

The specific choices that need to be made when picking an appropriate FF for a QM/MM simulation compromise on both, the origin of the reference data for the parametrization of the FF
and its specific functional form. With regard to the former, reference data can be acquired either from experiment or from QM calculations, both being costly. 

Here, we focus on molecular FFs for which the reference data for parametrization is obtained from QM calculations, which makes the training data generation agnostic with respect to the molecular composition, which is guaranteed by the QM fundament. We highlight parametrization strategies that can cope with large molecules, which is critical for QM/MM modeling.

\subsection*{Transferability versus System-Focused Parametrizations}

Universal (all-atom) FFs are parametrized based on \textit{ab initio} reference data and also on experimental data. For instance, the CHARMM force field and its polarizable variant are parametrized on Hartree-Fock/6-31G(d) interaction energies~\cite{Zhu2012} and on MP2/6-311G(d, p) reference data\cite{Lopes2013}, respectively.
Usually (and deliberately), they are not validated according to their ability to accurately reproduce a system's accurate structure or PES, but to reproduce, for example, bulk properties~\cite{Avula2021}, solvation free energies~\cite{Kashefolgheta2020}, or other experimental observables.\cite{Frohlking2020} 
Typically, transferable FFs assume that one potential energy expression can sufficiently well describe different, but chemically related compounds. 

This transferability assumption is often compromised for the following reasons: 
The parametrization and the accuracy assessment become difficult, if not impossible, when experimental data are scarce. Systematic improvement of universal FFs is very difficult, because the reference data for parametrization are not necessarily obtained from the system that is studied, and it is therefore difficult to detect regions of chemical space in which the FF fails to accurately model a system's specific PES. Moreover, parameters for nonstandard elements (e.g., transition metal atoms) are often lacking and fail to describe uncommon bonding patterns. 

The key insufficiency, however, is the FF ansatz itself, which is of limited transferability, because the physically relevant interactions of electrons and nuclei cannot be perfectly well convoluted into pair-wise interactions of united entities (such as atoms or united atoms) without loss of information and generality (even the inclusion of polarization effects builds upon approximate models in contrast to electronic structure models). As a consequence, FFs can only be applied within boundaries defined by the target property for which they were fitted (i.e., specific calculated properties or experimental observables).

Attempts have been made to alleviate inaccuracies of transferable FF parameters by system specific tuning~\cite{Frohlking2020, Zgarbova2015, Addicoat2014, Hudson2018, Giese2019}, and FFs have been parametrized systematically with respect to dense experimental data~\cite{Kashefolgheta2020} or QM reference data.\cite{Ercolesi1994}
Transferable FFs may still fail in certain situations; for instance, in the description of short-range repulsion interactions~\cite{Zgarbova2010}, whereas long-range noncovalent interactions are better described by mature transferable force fields.

By contrast, system-specific FFs attempt to build the parametrization on reference data derived mainly or exclusively from tailored QM calculations,\cite{Maurer2007, Grimme2014a, Hagler2015, Vanduyfhuys2015, Brunken2020} which is, by its very nature of the first-principles ansatz, available for molecules composed of any element of the periodic table. The resulting potential energy MM model then approximates the electronic PES of the reference system in the Born--Oppenheimer approximation. Such FFs are often called \textit{ab initio} FFs.  
QM-derived \textit{ab initio} FFs are, for instance, QMDFF~\cite{Grimme2014a} and QuickFF~\cite{Vanduyfhuys2015, Vanduyfhuys2018}. 
Their main obstacle is rooted in the computational expense of their parametrization, scaling at least with the cost of the applied QM method times the number of molecular structures for which these calculations are to be carried out. However, in view of the QM/MM approach, the choice of the same method for both the MM parametrization and the QM calculation leverages a consistency in the model construction in such a way that the entire model is built to approximate the same PES. This can be crucial because the compatibility of MM and QM methods is key for accurate QM/MM simulations, see, for example, the interactions with a standard water model, which can, depending on the QM method, distort energies significantly.\cite{Koenig2016}

FFs obtained by system-focused parametrization for a specific molecular system are, in general, not transferable. However, one may explore transferability  of results obtained for related chemical structures (e.g., in a $\mathrm{\Delta}$-ML approach that considers the specific differences of local molecular environments on top of a generalized FF, see, for example, Ref. \cite{Brunken2020}).

 Generality usually comes at the cost of a loss of accuracy in transferable FFs, whereas molecular-specific FFs suffer from computationally expensive parametrization. To alleviate the cost of parametrization, existing transferable FFs can be tuned according to additional reference calculations.\cite{Sami2021} Then, the usual lack of systematic improveability renders system-focused potential energy functions a superior choice. ML FFs can achieve transferability by on-the-fly extension of the training data set (we will turn our attention to ML FFs in Section \ref{sec:UQ}).

\subsection*{Functional Form and Parameter Evaluation}
FFs rely on simple analytic expressions composed of additive covalent and noncovalent contributions to the MM energy, 
\begin{equation}
	E_{\mathrm{MM}} = E_{\mathrm{cov}} + E_{\mathrm{noncov}} = E_\mathrm{r} + E_{\mathrm{\alpha}}  + E_{\mathrm{\psi}} + E_{\mathrm{\phi}} + E_{\mathrm{estat}} + E_{\mathrm{VdW}}, 
	\label{eq:MM_energy}
\end{equation}
for bonds $r$, angles $\alpha$,  dihedral angles $\psi$, improper dihedral angles $\phi$, and noncovalent interactions comprising electrostatic (estat) and van der Waals (VdW) interactions.
Additional corrections for polarization effects may be added, which we consider later. 

In Equation (\ref{eq:MM_energy}), the bonded terms only describe local interactions that can be modeled by harmonic oscillator (bonds, angles), double well (improper dihedral), and cosine-type (dihedral) potentials. 
These model potentials do not accurately describe anharmonic properties such as internal rotations with very low barriers and correlated dihedral potentials~\cite{Friederich2018}. However, various correction potentials have been devised, which solve these issues by additional terms in the bond angle terms\cite{Vanommeslaeghe2010}, dihedral terms\cite{Best2012}, and by applying Fourier expansions of the dihedral terms.\cite{Blondel1996}
However, the number of force constant parameters grows significantly upon consideration of anharmonic effects and complex functional forms often have limited compatibility to highly optimized MD software~\cite{Sami2021}. One way out of this dilemma is to correct for such effects a posteriori through on-the-fly $\mathrm{\Delta}$-ML.\cite{Brunken2020}

The nonbonded electrostatic and nonelectrostatic parameters, by contrast, model long-range interactions and are therefore computationally far more demanding for large structures. 
The van der Waals interaction term contains attractive and repulsive components and can be parametrized to reproduce experimental neat liquid densities and enthalpies of vaporization.\cite{Boulanger2018a} However, this parametrization can be tricky for hydrogen bonds, which are governed by dispersive and electrostatic effects. Hence, hydrogen bonds are sensitive to both the LJ parameters and the charges chosen for the electrostatic term (that can be derived by a plethora of different approaches~\cite{Mei2015, He2020}), which implies that both parameter sets must be mutually consistent.\cite{Riquelme2018}

Their functional form is often taken to be of Lennard--Jones (LJ) or related types~\cite{Buckingham1938, Halgren1992}, which suffer from the inability to describe (dynamic) polarization effects (see also Section \ref{sec:embedding}). Several groups investigated the direct derivation of LJ parameters from QM calculations; for example, by atoms-in-molecules electron density partitioning~\cite{Cole2016, Horton2019} or the exchange-hole dipole moment model~\cite{Mohebifar2017, Walters2018}, which can also be applied to obtain higher-order dispersion terms as shown by Visscher and coworkers.\cite{Visscher2019}

The electrostatic component of classical FFs is generally modeled as a Coulomb potential energy expression in terms of fixed atomic partial charges (which can be derived by direct fitting to the electrostatic potential~\cite{Chen2010}). This representation, however, fails to describe (i) anisotropic charge distributions, (ii) charge penetration effects (i.e., the overlap between charge distributions of atoms that approach one another),
and (iii) electrostatic polarization, that is, the change in charge distribution as a response to its environment.\cite{Baker2015} 

The explicit description of polarization effects has been applied in the context of polarizable FFs~\cite{Shi2013, Baker2015, Jing2019,Lin2020, Nochebuena2021}, 
for which two categories of polarization models exist: charge redistribution and charge flow. The latter describes polarization effects by the fluctuating charge model\cite{Rick1994, Rick1996, Cappelli2016}, in which atom-centered charges vary as a response to the environment during a simulation. As a charge-only framework, this approach employs a standard Coulomb potential and needs no further terms in the MM potential, but fails to describe anisotropic charge distribution within a molecule. To cure this deficiency, Giovannini et al. introduced the fluctuating-charge--fluctuating-dipole approach, in which both charges and higher-order multipoles can change in response to a surrounding electric field.\cite{Giovannini2019a}

In the charge redistribution framework, one can apply the induced dipole model or the Drude oscillator model. In the induced dipole model, each site is assigned a charge and a corresponding induced dipole. This is conceptually related to the Drude model~\cite{Boulanger2014, Lemkul2016, Lemkul2017}, where each atom is equipped with a charge pair, with one charge centered at an atom and the second charge assigned to a virtual Drude particle. 
This Drude particle is attached with a spring and can therefore induce a dipole moment.
Additional restraints that avoid large Drude-atom displacements are in place to avoid a polarization catastrophe~\cite{Huang2018}. Both methods require an iterative self-consistent field type solution, which makes them computationally more demanding. Note that modern FFs
not only account for polarizability, but also for anisotropic charge distribution through the inclusion of distributed multipoles instead of charges only\cite{Mao2016, Loco2016, Loco2019, Unke2017}, as, for example, in the AMOEBA FF.~\cite{Shi2013} For a detailed overview, we refer the reader to Refs. \cite{Baker2015, Jing2019}.

Polarizable FFs can be directly fitted to the continuous electron density; an example is the Gaussian Electrostatic Model (GEM).\cite{Cisneros2005} 
With respect to the functional form, a self-polarization term $E_\textrm{self}$ is then added to the total electrostatic term:
\begin{equation}
	E_{\textrm{estat}} = E_{\textrm{Coulomb}} + E_{\textrm{self}}.
\end{equation}
The additionally required parameters are, depending on the applied polarization model, the induced dipole moments, force constants of Drude atom displacement (note, however, that the Drude formalism and the induced dipole model can be mapped onto each other~\cite{Huang2017}), or electronegativities, hardnesses, and partial charges (which can be formulated in the QM framework).
Although atomic polarizabilities are no observables, they can be fitted to the ESP within response theory~\cite{Wang2016} or they can be derived from experimental data.\cite{Miller1978} 

In the QM/MM context, efficient implementations of polarizable QM/MM methods have been presented~\cite{Lambros2020, Curutchet2009, Loco2016, Caprasecca2015, Lipparini2019, Kratz2016, Loco2019, Reinholdt2020}  and have been applied, for instance, in studies of enzyme catalysis~\cite{Ganguly2017}, solvation effects~\cite{Dziedzic2016}, free energies of reaction~\cite{Sahoo2018}, electrocatalysis~\cite{Naserifar2021}, electronic~\cite{Loco2016, Olsen2010} and nuclear~\cite{Steinmann2017} spectroscopy, and QM/MM-MD.\cite{Hofer2018a, Loco2017}

For a given functional form of the MM potential, the parameters can be fitted to QM reference data and efforts have been made to parametrize both nonbonded~\cite{Vandenbrande2017, Cole2016, Allen2019} and bonded parameters~\cite{Allen2019}, for example, with the partial Hessian fitting approach.\cite{Wang2016a, Wang2018} 
Since the calculation of the Hessian matrix for the fitting of force constants is expensive, this limits the applicability of most QM parametrization techniques to medium-sized molecules or homogeneous MM environments (e.g., solvent molecules\cite{Visscher2018, Visscher2020, Asthana2013}). 
To parametrize condensed-phase systems of thousands of atoms, fragmentation strategies can be applied, in which the overall parameters are assembled from a set of reference calculations on smaller fragments of the nanoscopic system (as, for instance, in the SFAM FF~\cite{Brunken2020}).

Available, system-specific, QM-derived FFs are, for instance, the Quantum Mechanically Derived Force Field~\cite{Grimme2014a} (QMDFF), the SFAM FF\cite{Brunken2020}, the CUBE FF~\cite{Allen2019}, and the QuickFF force field~\cite{Vanduyfhuys2015, Vanduyfhuys2018}. While all FFs derive bonded parameters from QM calculations, they mainly differ in the degree to which empiricism is introduced in the nonbonded parameters. Also, not all parametrization strategies can be applied to large molecular systems. An overview of common QM-derived FFs is compiled in Table \ref{tab:qm_forcefields}. 

\begin{table}[h!]
    \centering
    \caption{Overview of system-specific QM-derived FFs. FFs marked with an asterisk augment classical transferable FFs by additional reference calculations. All bonded parameters are fitted to optimized geometries and molecular Hessians. Superscript $^{a)}$ denotes that QM torsional scans are additionally performed for parametrization of dihedral terms. Superscript $^{b)}$ refers to the application of fragmentation algorithms for transferability to larger molecular systems. Superscript $^{c)}$ denotes that these terms contain both system-specifically fitted and global parameters. ``No" marks that those parameters are not obtained from \textit{ab initio} reference data, but are set \textit{a priori} or taken from other FFs. All FFs are minimally assembled from the terms in Equation (\ref{eq:MM_energy}), additionally terms are colored in green, missing terms with respect to Equation (\ref{eq:MM_energy}) are colored in red. For details on notation, we refer to the original literature. }
    \label{tab:qm_forcefields}
\resizebox{\textwidth}{!}{\begin{tabular}{lccccc}
\hline
Force Field  & bonded from QM & nonbonded from QM &
polarizable & \begin{tabular}[c]{@{}c@{}} applicable to \\ large molecules \end{tabular} & available \\
\hline
\hline
QMDFF\cite{Grimme2014a} & \begin{tabular}[c]{@{}c@{}} $\color{Green} \mathrm{E_{r}^{13}}$ \\ yes$^{a)}$ \end{tabular}  &\begin{tabular}[c]{@{}c@{}}   $ \color{Green} \mathrm{E_{hb}, E_{xb}, E_{pol}}$ \\ $\mathrm{E_{estat}}$, $\mathrm{E_{hb}}$, $\mathrm{E_{xb}}$: yes$^{c)}$ \end{tabular}&  empirical & no & yes  \\
\hline
SFAM~\cite{Brunken2020} &\begin{tabular}[c]{@{}c@{}} yes \end{tabular} &\begin{tabular}[c]{@{}c@{}} $\color{Green} \mathrm{E_{hb}}$ \\  $\mathrm{E_{estat}}$: yes \end{tabular} & no & yes$^{b)}$ & \textsc{Swoose~\cite{Swoose2021}} \\
\hline
CUBE~\cite{Allen2019}  & \begin{tabular}[c]{@{}c@{}} $\color{Red} \mathrm{E_{\phi}}$ \\ yes$^{a)}$ \end{tabular}  & $E_{estat}$, $\mathrm{E_{VdW}}$: yes (AIM)\cite{Cole2016} & no &  $\thicksim$ 100 atoms & \begin{tabular}[c]{@{}l@{}}QUBEkit~\cite{Horton2019} \& \\ QUBEkit-pro~\cite{Nelson2021} \end{tabular} \\
\hline
Q-Force\cite{Sami2021}* & \begin{tabular}[c]{@{}c@{}} yes \end{tabular} & no   & no & yes $^{b)}$  & yes\\
\hline
Vilhena et al.\cite{Vilhena2021}  & \begin{tabular}[c]{@{}c@{}} yes$^{a)}$ \end{tabular}  & \begin{tabular}[c]{@{}c@{}} $\mathrm{E_{estat}}$: yes \\ $E_{VdW}$: yes$^{c)}$ \end{tabular}  & no & \begin{tabular}[c]{@{}l@{}}complex fluids\end{tabular} & yes \\
\hline
QuickFF\cite{Vanduyfhuys2015, Vanduyfhuys2018} &  \begin{tabular}[c]{@{}l@{}} yes$^{a)}$ \end{tabular}   & no & no & \begin{tabular}[c]{@{}c@{}} yes \\ target: MOFs \end{tabular}  & yes \\
\hline
\hline
\end{tabular}}
\end{table}

\subsection*{Automated Force Field Construction}
The parametrization of FF parameters follows a three-step process: After the selection of reference data, one needs to define an objective function and optimize the parameters to improve the value of this function. 
This workflow makes FF parametrization well suited for automated procedures~\cite{Jing2019} and multiple protocols have emerged that derive MM parameters from QM reference data in an iterative manner (see, e.g., Refs. \cite{Maurer2007, Barone2013, Schwörer2013, Wang2013, Grimme2014a, Wang2014, Prampolini2015, Vanduyfhuys2015, Prampolini2016, Allen2018, Cerezo2018, Fang2018, Horton2019, Visscher2019, Brunken2020, Vilhena2021,Seo2021, GreffdaSilveira2022}).
We choose to discuss only a few in some more detail for illustrative purposes:  Welsh et al.~\cite{Welsh2019} introduced the \textsc{CherryPicker} algorithm, which was designed for large biomolecules by combination of a fragmentation approach and a graph-based database search of already parametrized subentities. Furthermore, the \textsc{Joyce} and \textsc{Picky}
procedures for parametrization of intra- and intermolecular terms~\cite{Cacelli2012, Barone2013} are promising candidates for automated FF construction due to their high degree of flexibility with regard to the functional form and parameter origin. Originally applied only to small molecules, the \textsc{Picky} procedure was subsequently combined with 
a fragmentation approach and applied to complex fluids, for which the resulting FF outperformed both a transferable and a partially QM-derived FF (only bonded contributions).\cite{Vilhena2021} 
The fully automated parametrization pipeline for the construction of SFAM developed in our group couples a fragmentation ansatz~\cite{Brunken2020} to efficient parallelization of the reference data generation that are stored in a database. This pipeline is incorporated as the \textsc{Swoose} module~\cite{Swoose2021} into our \textsc{Scine} software project.

\section{Choice and Size of QM Region} \label{sec:QMRegion}
The choice and number of atoms to be included into the QM region embedded in the MM potential is a critical task: On the one hand, an optimal-size QM region must be capable of describing all relevant electronic correlations and processes that cannot be accurately modeled with the MM potential energy expression, and hence, require an electronic structure method. To evaluate which components of the entire model fulfill these criteria, is not trivial and requires extensive testing with multiple candidate QM regions of different size and composition. 
At the same time, the maximum size of the QM region is dictated by the affordability of the chosen electronic structure method, depending on the available computational resources and the type of simulation to be performed (i.e. single-point calculation vs. molecular dynamics). 

In search of a reliable QM-region size, the generation and selection of suitable candidate regions imposes several challenges: First, one requires systematic estimates on the accuracy of a calculated property with respect to a
very large reference region for which the property of interest is asymptotically converged. 
Second, one needs exploratory calculations on candidate models that are representative of
the conformational space, as the structural conformation may affect the accurate calculation of properties~\cite{Mehmood2020}. 
Unfortunately, this selection is often plagued by extensive manual input, its dependence on chemical intuition, and a lack of standardization. 
Consequently, algorithms must be able to find a minimum active region, which minimizes the difference of a property evaluated in the active region and a chosen reference.

Systematic determination of suitable QM regions has mainly been applied to protein structures, where chemical reactions occur at an active site (which can be identified and then incorporated into the QM region). Although the QM region has often been chosen based on intuition, we focus here on algorithms that allow for fully automated procedures, which are far less error-prone and save human time.
We review new-generation approaches for systematic QM region generation that target atom-economical choices to balance the computational effort and accuracy. 
Semiempirical QM/MM methods scale favorably with respect to the size of the QM region and even allow for very large QM regions, but their accuracy strongly depends on the specific semiempirical method and the application target. By contrast,
the standard approach of density functional theory often strikes a satisfactory accuracy--effort balance. 
Also, more accurate correlated methods have been applied successfully (see, e.g., Refs. \cite{Claeyssens2006} and  \cite{Bistoni2018}).

Various groups investigated and reported the convergence of properties as a function of a spherically growing QM region~\cite{Sumowski2009, Siegbahn2011, Siegbahn2011a, Flaig2012, Gomes2012, Isborn2012, Liao2013, Hartman2015, Harris2016, Kulik2016, Quesne2016, Himo2017}
The majority of studies applies distance-based criteria to construct spherical candidate regions around a central atom of choice. 
Distance-based (or shell-based\cite{Casalino2020, Torabifard2018a, Brunken2021}) criteria are easy to implement and fully automatizable.
However, it has also been shown that enlarging the QM region does not necessarily guarantee convergence of target properties.\cite{Fouda2016, Jindal2016, Sumner2013, Sumowski2009} 

The unfeasibility of simulations with large QM regions can be illustrated with a simple example of a QM/MM-MD simulation: 1 ns of simulation with a 0.2 fs time step amounts to a total process time on one central processing unit of approximately 5 years (depending on the computer architecture) under the assumption that a single QM calculation requires only 30 s of computer time, which applies only to very small QM systems and highly optimized QM software~\cite{Martins-Costa2017}, if one does not use semiempirical methods. 
Well parametrized semiempirical methods can be on a par or even better than conventional DFT methods in terms of accuracy of QM/MM simulations~\cite{Woodcock2007, Korth2011, Dral2016, Konig2016a}. Whereas the former are also orders of magnitude faster than the latter, the former are usually only available for a limited set of elements of the periodic table.  
Free-energy calculations require even hundreds of thousands of single-point energy calculations. Accordingly,  
many approaches have been devised to optimize scalability of QM/MM calculations~\cite{Schwinn2019, Schwinn2020, Huix-Rotllant2021, Reinholdt2021, Scheurer2021, Vennelakanti2022}, for example, on graphics processing unit architectures~\cite{Cruzeiro2021}, exploiting massive parallelization~\cite{Bolnykh2019, Olsen2019, Wang2015b, Suruzhon2021}, optimization of boundary schemes (e.g., the Electrostatic potential fitting (ESPF)\cite{Schwinn2019, Huix-Rotllant2021, Schwinn2020, Pan2021a}) and adjustment of FFs (cf., Section \ref{sec:force-fields}).

Moreover, the choice of multiple (potentially disconnected) QM regions~\cite{Torras2015, Borbulevych2018} instead of one large QM region can increase the overall efficiency of the simulation because each QM region can be calculated independently of the others in a parallel fashion.

Although polarizable embedding may allow for a decrease of the QM region size~\cite{Chen2021b, Flaig2012,Steinmann2014, Nabo2017}, because the environment's polarization response is included into the QM calculation, the accurate description of long-range charge transfer phenomena may still require QM regions of several hundreds of atoms.\cite{Kulik2018}

Many approaches have emerged to more systematically decide which subentities of a nanoscopic system should be included in the QM region.\cite{Karelina2017a, Hix2021, Brunken2021, Summers2021, Brandt2022} For instance, Kulik and coworkers~\cite{Karelina2017a} introduced the charge shift analysis (CSA) and the Fukui shift analysis (FSA) approaches. Changes in either the partial charges or the condensed Fukui functions of specific residues are evaluated as a response to modifications of the QM region. 

In the CSA approach, each residue in a protein is characterized by two residual partial charges computed as a simple sum of atomic charges in (i) the complete protein and (ii) the capped protein, in which the core site is altered or removed. Only if the residues' electronic properties are affected by the core site, the residue charge difference will exceed a certain cutoff. However, the partial charge difference needs to be calculated for very large reference QM regions (of about 1000 atoms) in which the electronic structure properties are asymptotically converged, hampering the application of this approach in high-throughput QM/MM calculations. Moreover, this metric depends on the type of partial charges and the specific electronic structure method chosen (such as the approximate exchange-correlation density functional selected), which needs to be capable of accurately describing charge transfer processes. 

In the FSA method, condensed Fukui functions of the core active site are computed in the presence and absence of each additional residue, one at a time. The magnitude in shift is evaluated with respect to the median Fukui values, with distant noninteracting residues serving as a reference. 
Because residues are added one at a time, FSA can be trivially parallelized. 
Even more, Fukui functions can be replaced by other first-principles heuristics such as electric dipole moments or bond critical points.  

These two approaches can be denoted ``top-down" and ``bottom-up", depending on whether the QM region is either iteratively shrinked from a large reference-region (as in CSA) or built up from a minimal core motif (as in FSA). 
Both, CSA and FSA, have been applied in various QM/MM studies and consistently yielded QM regions with quantitatively converged properties at only 50 percent size compared to reference calculations\cite{Karelina2017a}. Also, Chen et al. showed, at the example of proton-transfer reactions in solvents, that the application of charge-based schemes lowers the error in QM/MM reaction energies with respect to a full QM reference compared to distance-based schemes, irrespective of the choice of the FF.\cite{Chen2021b}

Both methodologies have in common that the nanoscopic structure needs to be categorized into (i) residues and (ii) a core active site, where candidate regions are incrementally enlarged by one residue at a time.
We proposed an alternative approach for systematic QM region determination, which is solely based on quantities obtained from the electronic energy and its derivatives with respect to the molecular coordinates in our QM/SFAM framework.\cite{Brunken2021} 

In QM/SFAM, candidate QM regions are generated around a central atom making use of the fragmentation approach proposed for the SFAM MM parametrization~\cite{Brunken2020}; this is then coupled to a stochastic element, where bonds are cut with some probability $p$. 
QM/SFAM allows for the fast generation of multiple candidate (and reference) regions within some upper and lower boundary in terms of molecular size set according to available computational resources. 
The suitability of a candidate QM region is subsequently evaluated by the mean absolute error of the force components on representative atoms in the QM regions with respect to the larger reference regions. 

\begin{figure}[ht!]
    \centering
    \includegraphics[width=1.0\textwidth]{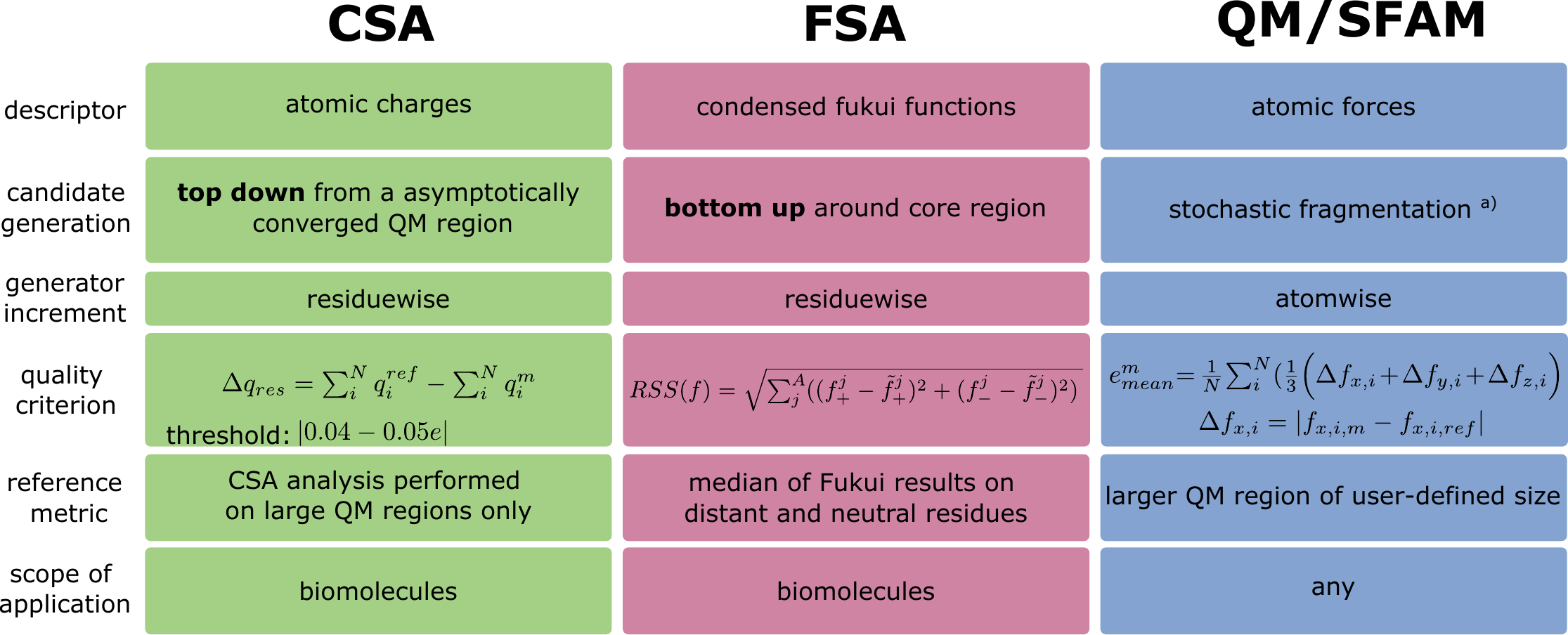}
    \caption{Comparison of three QM region selection algorithms: CSA~\cite{Karelina2017a}, FSA~\cite{Karelina2017a} and QM/SFAM.\cite{Brunken2021} In the CSA approach, ${\Delta q_{res}}$ denotes the charge shift for a specific residue defined as difference between the sum of the partial charges of the noncore residue in the complete protein ${q_i^{ref}}$ and the capped (mutated) protein ${q_i^{m}}$ (given in units of the elementary charge $e$). In the FSA approach, RSS(f) denotes the root sum squared difference in by residue Fukui functions $f_+$/$f_-$ of an active site residue $A$ with respect to the corresponding median $\tilde{f}_+$/$\tilde{f}_-$. In the QM/SFAM approach, ${e_{mean}^e}$ denotes the mean error of the force components of a QM/MM model $m$, calculated from the weighted sum ($N$ equals the number of atoms in the QM region) of the difference in force components $\Delta f_i$ for atom $i$ in $x$, $y$, and $z$ direction for the model ($f_{i, m}$) and the reference (${f_{i, ref}}$). 
    }
    \label{fig:qmRegions}
\end{figure}

QM/SFAM requires only one single-point gradient evaluation for each candidate and reference region and is, therefore, computationally comparatively efficient. 
As all candidate regions are independent of each other, all calculations trivially parallelize. Compared to CSA and FSA, QM/SFAM does not split the nanoscopic structure residue-wise, although cleavable bonds should be chosen with care for the type of system. 
In Figure \ref{fig:qmRegions}, we compare the CSA and FSA approach with QM/SFAM according to different criteria (the choice of descriptor, the strategy for candidate generation, the quality criterion for a candidate region and the corresponding reference, and the scope of application). 

We emphasize that the systematic determination of QM regions for nonbiomolecular structures (e.g., for heterogeneous catalysis on molecular surfaces) has received only little attention. In these cases, one can either combine a two-dimensional periodic QM approach with the QM/MM methodology in such a way that the QM/MM boundary does not break the periodicity~\cite{Hofer2015} or employ clusters of atoms~\cite{Sainna2021} as QM regions. The latter option was also applied in the study of covalent organic and metal organic framework structures, in which several unit clusters are taken as the QM region~\cite{Jones2020, Xu2017, Nakagaki2020, Yang2021}.

\section{Embedding and Handling of the QM/MM Interface}
\label{sec:embedding}
The QM/MM interaction energy (third term in Equation \ref{eq:qmmm-energy})  consists of covalent and noncovalent, that is, van der Waals and electrostatic interactions~\cite{Lin2006}:
\begin{equation}
E_{\mathrm{int}}^{\mathcal{Q} + \mathcal{E}} = E_{\mathrm{cov}}^{\mathcal{Q} + \mathcal{E}} +
E_{\mathrm{noncov}}^{\mathcal{Q} + \mathcal{E}} = E_{\mathrm{VdW}} ^{\mathcal{Q} + \mathcal{E}} + E_{\mathrm{cov}}^{\mathcal{Q} + \mathcal{E}} + E_{\mathrm{estat}}^{\mathcal{Q} + \mathcal{E}}. 
\label{eq:interaction_energy}
\end{equation}
These individual contributions can be treated on the MM level only (denoted mechanical embedding), incorporated into the QM Hamiltonian (then called electrostatic embedding or EE), or treated both quantum mechanically and classically (denoted polarizable embedding or PE). 
Note, however, that mechanical embedding is somewhat outdated as there is no polarization of the QM region by MM point charges and will therefore not be considered further in this Section. 

Very often, EE implicitly refers only to the electrostatic terms, following the assumption that they contribute most to the overall interaction energy. However, EE allows only for monodirectional polarization, that is, the QM region is polarized by the MM region but not vice versa. This lack of bidirectionality is cured in the polarizable embedding scheme. By definition, PE is a more consistent electrostatic scheme, where the subregions $\mathcal{Q}$ and $\mathcal{E}$
mutually polarize each other until an equilibrated
charge distribution is reached. In practice, this can be realized by coupling of the QM region to a polarizable FF (see Section \ref{sec:force-fields}). 
From a practical perspective, polarizable QM/MM is an iterative scheme in which the QM electron density is polarized by environment point charges and the polarizable atoms of the MM region respond to the electric field generated by the MM point charges and the QM electron density until self-consistency is reached. Such self-consistency iterations are expensive, but can be avoided by exploiting an extended-Lagrangian approach.\cite{Lamoureux2003}

Van der Waals contributions are generally modeled by parametrized functional forms independent of the QM density. The prime example is the LJ
potential, which combines a repulsive Pauli term and an attractive dispersive term. Pauli repulsion and long-range electron correlation must be postulated in a classical framework, but they are crucial even for systems in which electrostatic terms are dominating and should not be neglected in the EE framework.
One way to include nonelectrostatic effects is the effective fragment potential method\cite{Gordon2001, Gordon2013, Slipchenko2017b}, 
which, however, diverges from the FF concept and will therefore not be discussed in detail here. A completely different approach was developed by Giovannini et al.\cite{Giovannini2017, Giovannini2019a} who express repulsion based on an auxiliary density on the MM subsystem and quantum dispersion based on the Tkatchenko--Scheffler approach.

The contribution of covalent interactions $E_{\mathrm{cov}}^{\mathcal{Q} + \mathcal{E}}$ to Equation (\ref{eq:interaction_energy}) is nonzero only if covalent bonds cross the QM/MM boundary.  
This can be challenging as macromolecular structures exhibit different types of bonding patterns, ranging from covalent bonds to dative (between a metal and its ligand) and ionic (in crystals) interactions.

In this Section, we discuss different embedding methods 
defining Equation (\ref{eq:interaction_energy}) with a focus on the inclusion of explicit Pauli repulsion and dispersion into the QM/MM interaction energy expression in a mutually polarizable framework.

\subsection*{Contributions to the Interaction Energy I: Electrostatic Interactions}
Polarization of the QM region by MM point charges in EE can be achieved by augmenting the electronic ('$ele$') Hamiltonian of the QM region with an electrostatic potential of the MM charges (in Hartree atomic units used throughout):~\cite{Dohn2020} 
\begin{equation}
    \hat{H}_{\mathrm{ele, EE}}^{\mathcal{Q}} = \hat{H}_{\mathrm{ele}}^{\mathcal{Q}} -
    \sum_i^{N_{ele}^\mathcal{Q}} \sum_{A \in \mathcal{E}} \frac{q_A}{|\textbf{r}_i - \textbf{R}_A|} - \sum_{\alpha}^{N_{nuc}^\mathcal{Q}}
    \sum_{A \in \mathcal{E}} \frac{q_A Z_{\alpha}}{|\mathbf{R}_{\alpha} - \mathbf{R}_A|} 
	\label{eq:electrostatic_hamiltonian}
\end{equation}
with  the QM-internal Hamiltonian 
$\hat{H}_{ele}^{Q}$, 
the number of nuclei and electrons in the QM region $N_{nuc}^\mathcal{Q}$ and $N_{ele}^\mathcal{Q}$, respectively, the charge of MM atom $A$, $q_A$, the nuclear charge of nucleus $\alpha$ in $\mathcal{Q}$, $Z_{\alpha}$, 
and the Cartesian coordinates of electrons and nuclei $\mathbf{r}_i$ and $\mathbf{R}_A$, $\mathbf{R}_\alpha$, respectively. 
The required MM charges are readily available in all analytical FFs, but need to be chosen with care, because they may produce overpolarization in the QM region. 
This overpolarization is an artificial effect induced by the Coulombic interaction between the QM density and the MM point charges at small distances (i.e., close to the interface), where a quantum description of Pauli repulsion (which would be exerted by the MM electrons on the QM density in a full quantum system) is missing.

In metallic systems, where the effective atomic charges of the environment are zero, electrostatic embedding must be performed by the placement of effective core potentials on all atoms close to the QM region that mimic the presence of bulk around the QM atoms.\cite{Hofer2015} 
The number of additional one-electron terms in the QM Hamiltonian of Equation (\ref{eq:electrostatic_hamiltonian})
scales linearly with the number of MM atoms, which produces a computational bottleneck for large molecular systems. This computational overhead can be mitigated by truncation, i.e., by neglecting long-range
electrostatic interactions beyond some chosen cutoff. However, this can produce artifacts because long-range electrostatic interactions may become important for the description of condensed-phase processes.\cite{York1993, York1994, York1995} 

An alternative approach is to choose charges and (optionally) higher-order multipoles as a surrogate description of the continuous QM electron density.\cite{Olsen2015a, Beerepoot2016, Cardamone2014, Ferre2002} These multipoles must be fitted to reproduce the QM electrostatic potential arising from the correct QM density. It has also been suggested to approximate the continuous QM density through implicit solvent models~\cite{McCann2013, Acevedo2014}, MM-charge projection schemes~\cite{Pan2018, Pan2021a}, and Ewald-based methods.\cite{Giese2016, Kawashima2019}

Through combination of both continuous and surrogate schemes in a more advanced picture, the explicit QM density can be
employed in the short-range regime, while surrogate models for the QM density are chosen in the long-range regime~\cite{Fang2015, Kratz2016, Vasilevskaya2016a, Pan2018, Holden2019, Pan2021a, Kirsch2021}   
(see Ref. \cite{Pan2021a} for a detailed discussion). 

The MM contribution to the electrostatic embedding is purely parametric and static. 
At short range, a lack of Pauli repulsion can produce the electron-spillout problem, which is most prominent if the one-electron basis set in the quantum region extends far into the valence, and hence, into the MM region at the QM/MM interface. At small distances of QM and MM atoms, the QM electron density is overpolarized by the Coulombic interactions with positively charged MM atoms, whereas the Pauli repulsion in the electronic wave function induced by the (lacking) MM electrons is missing.
Mitigation of such inaccuracies usually requires extensive testing for a specific system under investigation, which cannot be routinely done in a black-box manner. The short-range Pauli repulsion can be approximately modeled by placement of pseudopotentials on close-lying MM atoms or by modified potentials for the Coulombic term (see also Ref \cite{Dohn2020} and references therein). 

PE has emerged as an advanced embedding scheme~\cite{Bondanza2020} that mitigates the inaccuracies arising from the static nature of electrostatic embedding.\cite{Lin2006, Illingworth2008, Ganguly2017} It was applied to describe charge penetration and anisotropy effects~\cite{Lin2006, Illingworth2008, Ganguly2017} and in the calculation of infrared spectra~\cite{Giovannini2019}, redox properties, and solvatochromic shifts of chromophores in solvents, and
biomolecules~\cite{Beerepoot2014, Tazhigulov2019, Ambrosetti2021} and in cases where the MM part experienced a strong polarization response, for example, upon electronic excitation in the QM region.\cite{Hagras2018a} 
We note that polarizable force fields often do not perform better than advanced fixed-charge force fields~\cite{Mohamed2016, Ganguly2017, Koenig2018}, which, however, may be simply taken as an indication that more development and testing will be required until the same degree of maturity is reached as for fixed-charge force fields.

In mutually polarizable QM/MM models, an additional term is added to the sum in Equation (\ref{eq:interaction_energy}), $ E_{\mathrm{pol}} ^{\mathcal{Q} + \mathcal{E}}$, corresponding to the polarization energy.\cite{Nochebuena2021} The incorporation of self-consistent mutual polarization lowers the computationally prohibitive demands of the electronic structure calculation, because the QM region does not need to be so large to cover 
all polarization effects.\cite{Daday2015, Nabo2017, Steinmann2014} For example, Loco and coworkers have shown that polarizable QM/MM models are less sensitive to the choice of the QM region.\cite{Loco2019}

To recover short-range nonelectrostatic interactions between the QM region and a (polarizable) MM region, three-layer QM/QM/MM schemes have been proposed.\cite{Olsen2015}
In such schemes, the nearest-neighbor 
environment is modeled through its (frozen) density in such a way that short-range Pauli repulsion can be recovered, and only distant MM atoms are described by multipole moments and polarizabilities. At the same time, polarization effects in the environment are described through classical polarizabilities, which are updated in a self-consistent manner as in the PE approach.

\subsection*{Beyond Polarizable Embedding: A Flexible QM/MM boundary}
\begin{figure}[h!]
    \centering
    \includegraphics[width=1.0\linewidth]{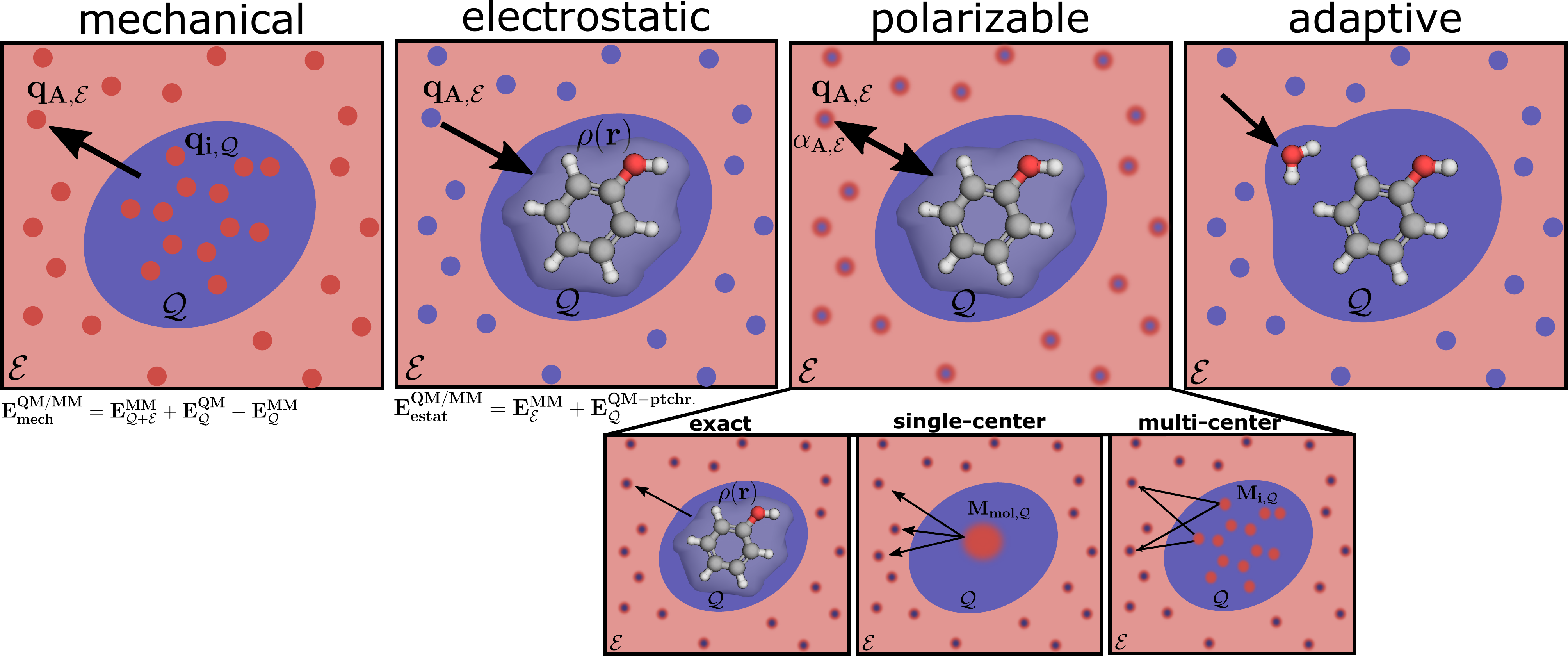}
    \caption{Schematic overview of QM/MM embedding models for the electrostatic contributions across the QM/MM boundary. Atom-centered charges in the MM sphere are depicted by circles, where the circle color refers to the regime at which electrostatic interactions are described (blue: QM regime, red: MM regime, blue/red: both). $\mathrm{\rho(r)}$ refers to the QM electron density and $\alpha$ denotes the additional assignment of polarizabilities to each atom-centered charge.
    $\mathbf{M_{mol/i}}$ denotes a molecular (i.e., single-center) or atom-centered (i.e., multi-center) multipole expansion for the  approximation of the continuous QM charge density (as, for example, implemented in the CFOUR/MiMiC software\cite{Kirsch2021}). The adaptive QM/MM embedding method is depicted in the electrostatic-embedding scheme (blue circles), but can, in principle, be combined with any other of the schemes mentioned (for an overview on different flavors of polarizable embedding, we refer to Ref. \cite{Reinholdt2021}).}
    \label{fig:embedding_models}
\end{figure}

Within a fully polarizable embedding approach, one may allow for dynamic exchange of atoms between the QM and MM regions during the course of a simulation; for example, if QM atoms are affected by diffusion and leave the QM region. This results in discontinuities in forces and energies and leads to the violation of energy conservation, which can be alleviated by introducing a transition layer~\cite{Duster2017}, data-driven many-body representations,\cite{Lambros2020} or machine-learned potentials.\cite{Zhang2018b}
This so-called adaptive QM/MM methodology has received considerable interest in the past decade.\cite{Park2012, Rowley2012, Pezeshki2011, Pezeshki2014a, Pezeshki2014, Waller2014, Watanabe2014, Zheng2016a, Dohm2017, Duster2018, Watanabe2019, Boereboom2016a, Pezeshki2015} 
It has mostly been applied to diffusion of solvent molecules across the interface, but also to proton transfer reactions.\cite{Duster2019} 
For metrics on how to reassign atoms to different regions, we refer the interested reader to Refs. \cite{Pezeshki2014a,Duster2017,Zheng2016a}. 
An overview of electrostatic embedding models is compiled in Figure \ref{fig:embedding_models}.

\subsection*{Contributions to the Interaction Energy II: Van der Waals and Bonded Interactions}
\label{subsec:VdW_and_bonded_interactions}
Pauli repulsion and dispersion contributions to the QM/MM potential energy expression 
captured by LJ-type potentials or similar parametrized analytical functional expressions at the MM level 
do not contribute to the electronic QM Hamiltonian.\cite{Dohn2018} In the EE framework, the static charge distribution in the MM subsystem is incapable of describing charge anisotropy and, hence, electronic repulsion, which renders the simplified  LJ potential energy expressions unavoidable~\cite{Dohn2020} and initiated the development of different density-dependent QM/MM van der Waals models within polarizable embedding frameworks.\cite{Smith2012a, Kuechler2015, Slipchenko2017, Giovannini2017, Curutchet2018, Gokcan2018, Rojas2018} 

This requires a reliable
formulation of contributions for dispersion and repulsion for the QM Hamiltonian. 
For example, Kuechler and coworkers designed the charge-dependent exchange and dispersion 
model~\cite{Kuechler2015} which 
introduces a scaled overlap model that is a function of the atomic charge. 
Several groups~\cite{Curutchet2018, Giovannini2017} employ modified Tkatchenko--Scheffler semiempirical van
der Waals schemes~\cite{Tkatchenko2009} for the dispersion contribution.
The density-dependent repulsion energy can be expressed 
as a standard Poisson integral.
The MM density can be approximated by Gaussian-type basis functions for ``fictitious" MM electrons, as shown by Giovannini et al.\cite{Giovannini2017} 
This approach has been successfully applied to a simulation of nicotine in aqueous solution~\cite{Giovannini2017}, the calculation of hyperfine coupling constants~\cite{Giovannini2019b} and noncovalent interaction energies.\cite{Curutchet2018} 
Slipchenko and coworkers have further developed the aforementioned effective fragment potential method to include
exchange-repulsion~\cite{Rojas2018} and dispersion~\cite{Slipchenko2017} into the QM Hamiltonian and showed that the resulting exchange-repulsion-corrected polarizable method improved the description of solvent-to-solute charge-transfer processes.\cite{Rojas2020} 

For the description of covalent bonds at the intersection of QM and MM regions, the canonical approach consists of placing a link atom, double-link atom~\cite{Das2002}, or tuned-link atom~\cite{Wu2018} at one end of the cut interface bond. 
Also, local self-consistent field~\cite{Thery1994, Ferre2002a, Monari2013},
generalized hybrid orbitals~\cite{Gao1998, Pu2005}, pseudo-bond~\cite{Zhang1999, Zhang2005}, frozen orbitals~\cite{Philipp1999, Kairys2000}, and Yin-Yang atoms~\cite{Shao2007} were developed to handle covalent bonds across the QM/MM interface.

\section{Uncertainty Quantification}
\label{sec:UQ}
Every result of a computational simulation requires the assignment of an individual, system-specific error in order to be meaningful by itself without relying on other information on reliability and accuracy~\cite{Reiher2021}. This process is called UQ in the field of computational science. In principle, any physicochemical property model must be subjected to a UQ analysis,~\cite{Proppe2016, Simm2017} which has made considerable progress in recent years (see, for instance, Refs. \cite{Mortensen2005, Simm2016, Proppe2016,Pernot2017, Pernot2017a, Pernot2018, Simm2018, Proppe2019, Proppe2019a, Pernot2022}).
UQ in multiscale models has also been recognized as a key issue.\cite{Chernatynskiy2013, Ye2021} 
The necessity of careful error estimation is particularly pressing in view of the manifold of sources of errors in QM/MM simulations.

Errors are generally grouped into epistemic (or systematic) and aleatoric (or statistic) errors:  
Epistemic uncertainty in QM/MM simulations is induced by inadequacy of the chosen MM model, errors in the underlying QM method applied to the core subsystem, and the incomplete approximation of interaction energies (e.g., the neglect of Pauli repulsion). 
The main contributions to the uncertainty in MM are (i) the selected functional forms of the FF and (ii) the choice and quality of reference data.
Even if the error in the potential may contribute most to the overall epistemic uncertainty, the electronic structure method applied to the QM region is affected by an error that, albeit small, is unknown for a specific structure of interest.\cite{Reiher2021} Aleatoric uncertainty, by contrast, arises from the randomness of sampling from a distribution. The uncertainties of individual structure configurations propagate into errors of thermodynamic averages~\cite{Imbalzano2021}, and the choice of the starting structure in short simulations may lead to strong biases.\cite{Leitgeb2005, Hahn2020}
This can be alleviated with ensemble methods in which a huge number of simulations are run, where one can extract statistically significant results from the ensemble of simulations.\cite{Ryde2017, Wan2021} 

Developments in the field of molecular simulation have achieved high reliability (as is evident from the previous Sections)
and many best-practice protocols have been developed for aleatoric error control.\cite{Olsson2017, Gapsys2020} 
The impact of systematic errors on the overall uncertainty, however, requires more attention because it constitutes the major part of the final uncertainty~\cite{Cailliez2011,Messerly2017, Yildirim2020}. 
Hence, epistemic uncertainty limits the reliability of the outcome even in the hypothetical case of infinite sampling and perfect structural models.\cite{Yildirim2020} 

\subsection*{Epistemic Error Sources in Multiscale Models}
Various studies involving sensitivity analysis~\cite{Moore2017, Vassaux2021} show that property prediction is highly dependent on the choice of the FF parameters.\cite{Rocklin2013a, Best2008, Lindorff-Larsen2012, Messerly2018, Cailliez2020a}
Extended regions of parameter space can fit the reference data equally well and even if model training is insensitive to the choice of parameters, this must not hold true for predictive calculations. 
Insufficiency in the functional form and in the quality of parametrization of a classical potential yield nontransferable FFs. 
For instance, QM-derived FFs are mostly fitted to properties resulting from QM calculations performed at minimum-energy configurations. 
Hence, reliable simulations for structure configurations that are not represented in the parametrization data are challenging.

The faithfulness of an MM potential can be assessed either by the generation of accurate QM reference data\cite{Kaminski2001, Morgado2010, Kanal2018, Koenig2020} (which can become computationally prohibitive for large structures), or through Bayesian schemes, which have been applied successfully for the UQ of FF parameters.\cite{Rizzi2012, Cailliez2020a}
Bayes' rule allows for automated and statistically motivated update of parameters given new data.
For example, LJ parametrizations have been the target of Bayesian UQ due to its simplicity.\cite{Cailliez2011, Pernot2017, Angelikopoulos2012, Messerly2017, Wu2016,Weymuth2018,Proppe2019a} 
A disadvantage is that Bayesian schemes may feature a computationally demanding calculation of the error. 

Since the general purpose of an atomistic or QM/MM simulation is the exploration of unseen regions of the PES, frequent visiting of new configurations is intentional, which requires (i) metrics to identify  regions in which the model prediction is insufficient and (ii) algorithms that deliberately visit such regions without extensive sampling. 
Epistemic uncertainties in the MM model can rely on Bayesian inference rules. 
Bayesian schemes, such as Gaussian process regression, yield a probability distribution over quantities of interest. 
Then, UQ is readily achievable through the available confidence intervals. 
In combination with the reliability of such ML models, many of them based on Gaussian processes~\cite{Deringer2021, Glielmo2017, Glielmo2018, Glielmo2020, Bartok2018}, ML-based MM modeling \cite{Behler2007} has been established as a new approach for QM/MM models in the past decade.

\subsection*{The Potential of ML in Force Field Construction}
ML potentials have recently emerged as flexible and tunable interatomic potentials that can provide accuracy of first-principles methods at a fraction of the computational cost and offer, in addition, UQ. 
QM/MM simulations that employ ML potentials offer several advantages, similar to those of system-focused analytical \textit{ab initio} potentials, compared to universal, transferable force fields:
\begin{itemize}
	\item The mathematical definition of a ML potential is, in principle, universally valid, because the potential does not explicitly rely on element-specific or rigid analytical expressions. By virtue of its ansatz, an ML potential is agnostic with regard to the molecular composition and only depends on the training data used for parametrization. This mitigates the need to choose one specific readily available FF, which usually requires expert knowledge or extensive testing. Note that ML potentials come at the cost of extensive testing, because the mathematical expression of an ML potential is not derived from physical criteria and at the same time very flexible, which can result in heavily oscillating functions that connect the training data. 
	\item The parametrization of an ML potential can easily be automated. 
	\item If the ML potential is trained on QM reference data obtained with the same method that is applied to the QM region, the QM/MM simulation will refer to the same consistent PES in both the QM and MM region. 
	\item ML potentials can be systematically improved on the fly by integration of additional reference data, for instance, if regions exhibiting high prediction uncertainty are visited during the simulation.
\end{itemize} 
Note that in contrast to analytical FFs, ML potentials are not transferable to any configuration on the PES. This again originates from the lack of physical basis of the functional form, which extrapolates in regions of the PES that were unseen in the training data, where it adopts somewhat arbitrary values. 
The details on how to set up ML potentials has been discussed by various groups and is out of scope of this review (see, e.g.,  Refs. ~\citenum{Behler2007, Behler2015, Artrith2016, Khorshidi2016, Botu2017, Chmiela2017, Glielmo2017, Smith2017, Bartok2018, Glielmo2018, Deringer2019, Riniker2019, Behler2021, Friederich2021, Miksch2021, Unke2021}).

ML models suffer somewhat from a lack of a theoretical base model for the predicted phenomena. It may be difficult to reliably estimate whether the visited region on a PES corresponds to an extrapolation regime with respect to the training set.\cite{Imbalzano2021}
Even more, accurate learning requires thousands of QM derived reference points in a carefully designed training set, which makes the training process computationally expensive. This supported the development of so called active-learning strategies~\cite{Shapeev2020, Zhang2019}, where an initial potential is iteratively improved by adding new reference data corresponding to regions of high uncertainty to the training set on the fly.\cite{Li2015, Amabilino2019} 
This method was applied to FF training in an MD simulation.\cite{Vandermause2020} 
A different approach to error estimation has been based on the generation of  ensemble models~\cite{Deringer2021}, where one or multiple ML models are fit to randomly chosen subsets of the training set~\cite{Peterson2017a, Smith2018, Musil2019, Behler2014, Xiao2018, Novikov2019, Zhang2019} and the spread in the prediction is a measure for the uncertainty of the ML potential. 

Dropout neural network (NN) potentials, by contrast, provide Bayesian error estimates on a significantly reduced timescale compared to ML potential sampling.\cite{Gal2016, Gal2016a, Wen2020} 
Imbalzano and coworkers~\cite{Imbalzano2021} suggested a hybrid-UQ approach also based on a committee model error estimation scheme: For single-point uncertainty estimation, they employ $\mathrm{\Delta}$-ML for the interpolation between a less accurate potential to a more accurate model. 
As this error propagates into the thermodynamic average over a trajectory, reweighting schemes are applied on the fly. 

\subsection*{The Potential for ML in Hybrid QM/MM Calculations}
ML algorithms have gained attention in the QM/MM community beyond the use of the ML potentials. 
One advantage, for instance, is the improved simplicity in the description of the interaction energies of the subsystems $\mathcal{Q}$ and $\mathcal{E}$. 
If the ML model is retrained on each performed QM calculation, large discontinuities between the QM and ML description of the boundary interactions will cancel out (for details, see Ref. \cite{Zhang2018b} and references therein).
Consequently, adaptive QM/MM can be employed without the use of an interpolative transition layer, because the discontinuity in energies and forces vanishes upon retraining of the ML potential.\cite{Zhang2018b} To avoid boundary problems, Lier et al.~\cite{Lier2022}  introduced a NN approach in which the NN is applied to describe interactions between a buffer region and both the quantum and classical subsystems. The buffer region itself is polarized by the quantum density.

\textit{Ab initio} QM/MM calculation are computationally expensive. Semiempirical QM/MM calculations are several orders of magnitude faster, but their accuracy is largely dependent on the quality of the model parameters, for which systematic error estimation is difficult. 
Recent approaches apply NN representations~\cite{Shen2016, Wu2017, Shen2018a, Gastegger2021} and other ML schemes~\cite{Zhou2014, Wu2017, Shen2018a, Zeng2021a, Zheng2021} to reduce the computational cost of the QM calculation in the QM/MM frameworks, either by learning the \textit{ab initio}-QM/MM PES from the semiempirical QM/MM PES (denoted $\mathrm{\Delta}$-ML, which goes back to work of von Lilienfeld and coworkers~\cite{Ramakrishnan2015}) or by directly learning the \textit{ab initio} PES~\cite{Behler2014, Behler2015, Botu2017, Chmiela2017, Smith2017}, the electron density,\cite{Fabrizio2019} or the wave function~\cite{Qiao2020} of the QM subsystem. 
For instance, B\"oselt and coworkers applied a NN representation of the QM region coupled to a $\mathrm{\Delta}$-ML approach.\cite{Boselt2021} 
Their resulting (QM)ML/MM model reached the accuracy of a density functional theory reference in the calculation of energies and
gradients. 

Recently, Brandt and Jacob~\cite{Brandt2022} investigated the uncertainty in QM/MM reaction energies with regard to the choice of the QM region and the empirical parameters incorporated in the force field (in this case, the MM point charges): These authors evaluated the derivative of a QM/MM energy calculation with respect to variations in the MM point charges, which was exploited to systematically incorporate atoms into the QM region in order to minimize this uncertainty.

\section{Conclusions}
Despite the simple concept of QM/MM modeling to embed electronic structure methods in a scaffold rooted in classical mechanics at the level of energies, it presents several challenges in the model generation. These challenges threaten reproducibility, the estimation of the uncertainty of calculated data, and automation efforts. Decades of research in this field have delivered diverse best-practice protocols and state-of-the-art QM/MM approaches to target such deficiencies, which we reviewed in this work.  We identify four components critical for the construction of reliable hybrid models, that should be carefully considered in any QM/MM modeling study:

\begin{itemize}

\item QM/MM model construction can and should be automated. Since this has been accomplished only for some aspects of the QM/MM model, a full-fledged black-box-type QM/MM model-construction engine is desirable. 
\item Atom-economically embedded QM regions can be well determined and should therefore be selected by first-principles criteria. 
\item MM potentials should be parametrized in a system-focused manner with reference data obtained from an electronic structure method. By that, both embedded methodologies adopt high compatibility among each other with regard to good phase-space overlap in order to be structure-faithful. Moreover, reference data can be produced on demand. 
Transferable information for generalized models can be systematically identified in a second step.
\item Rigorous UQ is needed to make QM/MM results reliable.
Especially, ML techniques can be efficient for on-the-fly refinement of
FF parameters. However, this requires well-calibrated re-weighting schemes in simulations, as a change of the FF parameters obviously directly affects the microstate energies.
\end{itemize}

These criteria define an experimental-data-free bottom-up approach that is a viable way to address any nanoscale atomistic structure, and that is therefore worthwhile pursuing.

\section*{Funding Information}
The authors gratefully acknowledge financial support through SNF project no. 200021\_182400.
Moreover, Katja-Sophia Csizi was supported through SNF grant no. 200021\_172950-1 awarded to PD Dr. Thomas Hofstetter.

\section*{Conflict of Interest}
The authors declare that there is no conflict of interest.


\begin{thebibliography}{100}
\bibitem{Lin2006}
Lin,~H.;\ \ Truhlar,~D.~G.  {QM/MM: what have we learned, where are we, and
  where do we go from here?},  \textit{Theor. Chem. Acc.} \textbf{2007,}
  \textsl{117,} 185.
\bibitem{Duster2017}
Duster,~A.~W.;\ \ Wang,~C.~H.;\ \ Garza,~C.~M.;\ \ Miller,~D.~E.;\ \ Lin,~H.
  {Adaptive quantum/molecular mechanics: what have we learned, where are we,
  and where do we go from here?},  \textit{Wiley Interdiscip. Rev. Comput. Mol.
  Sci.} \textbf{2017,} \textsl{7,} 1-21.
\end{thebibliography}

\providecommand{\refin}[1]{\\ \textbf{Referenced in:} #1}

\end{document}